%% file: simplification_IT_final.tex
\documentclass{IEEEtran}
\usepackage{amsmath,amsfonts, amssymb,pst-all}
\usepackage[dvips]{graphicx}
\newtheorem{theorem}{Theorem}
\newtheorem{lemma}{Lemma}
\newtheorem{remark}{Remark}
\newtheorem{proposition}{Proposition}

\def\EE{\mathbb E}

\def\SNR{\text{SNR}}

\def\lp{\left(}
\def\rp{\right)}
\def\lb{\left[}
\def\rb{\right]}
\def\l{\Lambda}
\def\lbar{\overline{\Lambda}}
\def\N{ \left[ N \right] }
\def\n{ \left[ N \right] }
\def\wbar{\omega}

\def\w{\omega}
\def\g{\Gamma}
\newcommand{\set}[1]{\left[ {#1} \right]}

\begin{document}
\title{Wireless Network Simplification:\\ the Gaussian $N$-Relay Diamond Network}
\author{Caner Nazaroglu, Ayfer \"Ozg\"ur,~\IEEEmembership{Member,~IEEE}, and Christina Fragouli,~\IEEEmembership{Member,~IEEE}
\thanks{C. Nazaroglu is with the University of Chicago, Department of Physics, 5720 South Ellis Avenue, Chicago IL 60637, USA (e-mail:cnazaroglu@uchicago.edu). A.~\"Ozg\"ur is with Stanford University, Department of Electrical Engineering, 350 Serra Mall, Room 205 Stanford, California 94305-9510, USA (e-mail: aozgur@stanford.edu). C. Fragouli is with the Ecole Polytechnique F\'ed\'erale de Lausanne, Facult\'e Informatique et Communications, Building BC, Station 14, CH - 1015 Lausanne, Switzerland (e-mail: christina.fragouli@epfl.ch). This paper was presented in part at the IEEE Int. Symposium on Information Theory (ISIT), St Petersburg, July 2011.}
}
\maketitle
\begin{abstract}

We consider the Gaussian $N$-relay diamond network, where a source wants to communicate to a destination node through a layer of $N$-relay nodes. We investigate the following question: what fraction of the capacity can we maintain by using only $k$ out of the $N$ available relays? We show that independent of the channel configurations and the operating $\SNR$, we can always find a subset of $k$ relays which alone provide a rate $\frac{k}{k+1}\bar{C}-G$, where $\bar{C}$ is the information theoretic cutset upper bound on the capacity of the whole network and $G$ is a constant that depends only on $N$ and $k$ (logarithmic in $N$ and linear in $k$). In particular, for $k=1$, this means that half of the capacity of any $N$-relay diamond network can be approximately achieved by routing information over a single relay.  We also show that this fraction is tight: there are configurations of the $N$-relay diamond network where every subset of $k$ relays alone can at most provide approximately a fraction $\frac{k}{k+1}
$ of the total capacity. These high-capacity $k$-relay subnetworks can be also discovered efficiently. We propose an algorithm  that computes a constant gap approximation to the capacity of the Gaussian $N$-relay diamond network in $O(N\log N)$ running time and discovers a high-capacity $k$-relay subnetwork  in $O(kN)$ running time.

This result also provides a new approximation to the capacity of the Gaussian $N$-relay diamond network which is hybrid in nature: it has both multiplicative and additive gaps. In the intermediate SNR regime, this hybrid approximation is tighter than existing purely additive or purely multiplicative approximations to the capacity of this network.
\end{abstract}

\section{Introduction}

Consider a source connected to a destination through a network of wireless relays  arranged in an arbitrary topology.  There are several ways to use this network. For example, we can route the information from the source to the destination  over a single path, using point-to-point connections.
Or, following an information theoretic approach, we can seek to optimally utilize all the available relays  to achieve the network
capacity, the largest end-to-end communication rate this network can support. Clearly the first approach has lower complexity and uses fewer resources of the network, while the second can potentially achieve much higher throughput. In this paper, we aim to understand the fundamental trade-off between using fewer relays and achieving larger rates, and perhaps the possibility of having both at the same time. We ask the following question: can we achieve (a good part of) the capacity of a wireless network by using only a (small) subset of (perhaps a large number of) available relay nodes?

Traditionally, network information theory  aims to characterize the best  end-to-end communication rate we can achieve in a network, without providing any understanding of the importance of each relay for achieving this rate \cite{EGC79, KMY06, KGG05, AvDigTse09, LKGC09}. However, in order to design simple and efficient communication architectures for wireless networks, apart from knowing the capacity of a large network, it may be even more useful to know what is the largest rate we can achieve by  using only a given number of the relays. We may want to know how this rate increases if we allow for more relays; how it compares to the capacity of the network; and how to efficiently discover the subset of relays providing the largest capacity.

Such an understanding can help with the design of more energy efficient communication protocols, that better utilize the limited wireless resources.
For example, relay nodes that contribute marginally to capacity can be shut down to save battery life. Alternatively, different parts of the network,  can be activated one at a time for maximal power efficiency. In a network with multiple information flows, knowing how much different relay groups  contribute to the throughput of each flow, can allow for an informed allocation of the relays across different flows.

\begin{figure}[t]
\input{N-diamond.tex}
\caption{The Gaussian $N$-relay diamond network. The source is connected to the relays through a broadcast channel, while the relays are connected to the destination through a multiple-access channel.}
\label{fig:diamondN}
\end{figure}
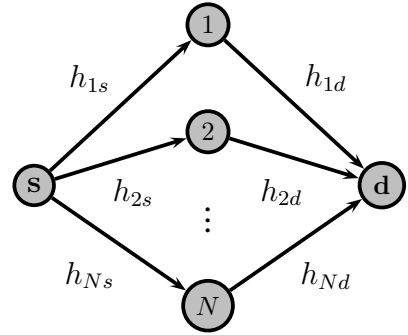

As a first step in this direction, in this paper, we consider a source that communicates to a destination over the Gaussian N-relay diamond network depicted in Fig.~\ref{fig:diamondN}. This is a two-stage network, where the source  node is connected to $N$ relays through a broadcast channel and the relays are connected to the destination through a multiple-access channel.  We ask, what fraction of the capacity we can achieve by using only $k$ out of the $N$ relays (for example, if we route the information between the source node and the destination over a single relay).

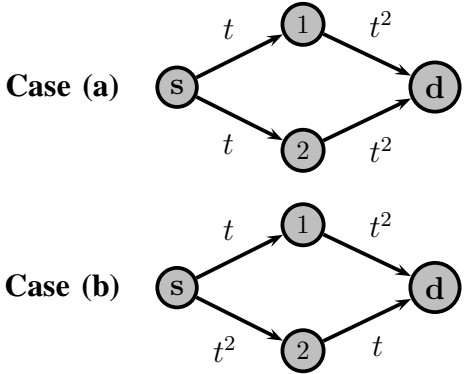
\begin{figure}[t]
\input{diamond.tex}
\caption{Two instantiations of a diamond relay network.}
\label{fig:diamond}
\end{figure}

The fraction of the capacity we can get with $k$ relays naturally depends on the channel gains. Indeed, consider for example the case where $N=2$, the  diamond  network, and the example in Fig.~\ref{fig:diamond}. For the identical channel gains in Fig.~\ref{fig:diamond}(a) we can show that  the communication rate achieved using only one of the relays  is only $1$ bit/s/Hz away from the cut-set upper bound on the capacity of the network; while for the anti-symmetrical channel gains as in Fig.~\ref{fig:diamond}(b) using only one of the relays  achieves (within 1 bit/s/Hz) only \emph{half} of the  cutset upper bound on the capacity of the network. 

To avoid channel-specific results, we  can try to provide worst-case guarantees that hold universally for all possible channel gains. For example,  is it possible that in $2$-relay networks, we can always find a single relay to use  and still achieve \emph{half} of the capacity of the diamond network within 1 bit/s/Hz (as  was the case for the two examples in Fig.~\ref{fig:diamond}). We prove in this paper that this is indeed always the case. In fact, we show that even if we have an arbitrary number $N$ of relays, we can remove all but one of them  and still achieve approximately half of the capacity of the whole network.

Our main result  is to show that in every Gaussian $N$-relay diamond network, there exist a $k$-relay sub-network whose capacity $C_k$ satisfies
\begin{align}
C_k&\,\geq\, \frac{k}{k+1}\, \overline{C}\,-1.3k- G\label{eq:mainres}
\end{align}
where $\overline{C}$ is the cut-set upper bound on the capacity of the $N$-relay diamond network and
$
G=\max\left(3\log N-\log\frac{27}{4}, 2\log N\right)
$
is a universal constant independent of the channel gains and the operating SNR. Intuitively, this holds because if all $k$-relay subnetworks have small capacity, the capacity of the whole network cannot be too large. As $k$ increases, the difference between the capacity of the best $k$-relay subnetwork and that of the whole network naturally decreases. The surprising  outcome here is that the fraction of the capacity we can get with $k$ relays is independent of the number of available relay nodes $N$. Moreover, it increases quite quickly with $k$: in the high-capacity regime, we can get at least half-the capacity of every $N$-relay diamond network by simply routing information over the best relay, using $2$ relays we achieve a fraction of $2/3$, etc.

We also show that the lower bound  in \eqref{eq:mainres} is tight in the multiplicative fraction, i.e.,
it is possible to find
$N$-relay diamond  networks where the capacity of \emph{every} $k$-relay sub-diamond network is at most 
\begin{equation}\label{eq:mainres2}
C_k\leq \frac{k}{k+1} C + G^\prime,
\end{equation} 
where $C$ is the capacity of the whole network and $G^\prime$ is a constant linear in $k$ and independent of everything else. For the case $k=1$ and $N=2$, one such example is case (b) in Fig.~\ref{fig:diamond}.

We prove the result  \eqref{eq:mainres} in two steps. We first show that in every Gaussian $N$-Relay diamond network, there exists a subset of $k$-relay nodes such that the information-theoretic cut-set upper bound on the capacity of this $k$-relay sub-network is larger than $\frac{k}{k+1} \overline{C}-G$; i.e., this step only involves the cut-set upper bounds on the capacities of the corresponding networks. We then use the compress-and-forward type of strategies in \cite{AvDigTse09,OD10,LKGC09}, over this $k$-relay sub-network. These strategies are known to achieve the cut-set upper bound on the capacity of any arbitrary Gaussian relay network within a gap that is linear in the number of relay nodes utilized. In particular, the result of \cite{LKGC09} implies that we can achieve the cut-set upper bound on the capacity of the $k$-relay network within $1.3k$ bits/s/Hz. Combining these two steps yields \eqref{eq:mainres}.

An alternative relaying strategy that is often considered for the $N$-relay diamond network in the literature is amplify-and-forward \cite{shashi, maric, ND10}. For example, \cite{ND10} shows that amplify-and-forward at the relays can achieve the cutset upper bound on the capacity of the $N$-relay diamond network within $3.6$ bits/s/Hz when all channel gains in the first and the second stages are equal. To show that this approximate optimality of amplify-and-forward is only limited to the the case of equal channel gains, we show that the rate achieved by this strategy is approximately equal to the capacity of the best relay alone in any arbitrary $N$-relay diamond network. More precisely, we show that
$$
C_{AF}\leq C_1 + 2 \log N
$$
where $C_{AF}$ is the best rate achievable with amplify-and-forward at the $N$ relays, and $C_1$ is the rate achieved by using only the best relay (say, with a decode-and-forward strategy) while keeping the rest of the relays silent. This result says that amplify-and-forward with the $N$ relays can at most provide a beamforming gain, bounded by $2\log N$, over the best relay. Since our result in \eqref{eq:mainres2} shows that there  are configurations of $N$-relay diamond networks where the best relay alone can at most provide approximately half the capacity of the whole network, the two results together imply that amplify-and-forward can be limited to approximately half the capacity of the network in certain configurations. This implies that amplify-and-forward fails to provide constant gap approximations for the capacity of the $N$-relay diamond network, such as those provide by the compress-and-forward type of strategies in \cite{AvDigTse09,OD10,LKGC09,flows} or partial-decode-and-forward in \cite{bobbie}.

Finally, a natural question given our existence result in \eqref{eq:mainres} is whether such high-capacity $k$-relay subnetworks can be discovered efficiently. Our existence proof naturally suggests an algorithm for discovering such networks in $O(kN)$ running time given the cutset upper bound $\bar{C}$ and the configuration of the $N$-relay diamond network. However, a direct computation of $\bar{C}$ itself requires evaluating the cut capacity over exponentially many cuts. \cite{parvaresh} shows that the problem of computing a constant gap approximation to $\bar{C}$ can be casted as a minimization of a submodular function and solved in $O(N^5\alpha+N^6)$ running time using state-of-the-art algorithms for submodular function minimization, where $\alpha$ is the time it takes to compute the value of a single cut which is typically polynomial in $N$. Our work reveals that  information flow in wireless networks has much more structure than mere submodularity. We show that the combinatorial structure that allows 
us to obtain the simplification result in \eqref{eq:mainres} can be also used to devise an algorithm to compute a constant gap approximation to the cutset upper bound on the capacity of the $N$-relay diamond network in $O(N\log N)$ time. The properties of wireless information flow beyond submodularity are further exploited in \cite{Naz2} where Non-Shannon properties of Gaussian random variables are used to obtain simplification results for the $N$-relay diamond network with multiple antennas.

\section{ Related Work  and Positioning}\label{sec:related}
Two lines of work have previously looked at a form of network simplification for wireless networks. First, relay selection techniques  in \cite{rel1,rel7,rel9}, design practical algorithms that allow to select the best single relay in an N-relay diamond network, and show that such algorithms provide cooperative diversity. These works look only at maintaining diversity and not  capacity. Second, work in \cite{ND10,rel2,rel4,rel8} looks at selecting a subset of the best relays when restricted to utilize an amplify and forward strategy. Our work differs  in that we do not restrict our attention to specific strategies (or a single relay) but instead provide universal capacity results for arbitrary strategies.

Our result can also be regarded as a new approximation to the capacity $C$ of the  Gaussian $N$-Relay diamond network. We show that
\begin{equation}\label{approx1}
 \frac{k}{k+1} \overline{C}\,-1.3k\,-\frac{k}{k+1}G\, \leq C\leq \overline{C}\quad \forall k,\,1\leq k\leq N-1,
\end{equation} where  $\overline{C}$ denotes the cut-set upper bound. The best of the earlier approximation results in \cite{AvDigTse09, OD10, LKGC09} yield 
\begin{equation}\label{ADT}
 \overline{C}-1.3N \leq C\leq \overline{C}. 
\end{equation}
for the $N$-Relay diamond network. 

The lower bound we provide in \eqref{approx1} is tighter than \eqref{ADT} in the intermediate SNR regime and when $N$ is large. The auxiliary parameter $k$ in \eqref{approx1} allows to optimize this lower bound as a function of $C$ and $N$. When $N$ is large, choosing a small $k$ reduces the additive gap from $O(N)$ in \eqref{ADT} to $O(\log N)$. This improvement in the additive gap can be more important than the $\frac{1}{k+1}\bar{C}$ loss due to the multiplicative gap when $\bar{C}$ (and therefore $C$) is not too large, overall yielding a tighter lower bound than \eqref{ADT}. When $C$ is large and $N$ is small increasing $k$ to $N$ reduces \eqref{approx1} to \eqref{ADT}. This approach suggests a new approximation philosophy to the capacity of wireless networks where multiplicative and additive gaps to the cutset upper bound are allowed simultaneously and are traded through an auxiliary parameter (in our case $k$). Earlier works in the literature have either aimed to characterize the capacity within an 
additive gap by allowing no multiplicative gap \cite{AvDigTse09, OD10}, or vice-a-versa \cite{ND10}. These purely additive or multiplicative capacity approximations are relevant in the high or the low SNR regimes respectively, while a hybrid approximation can be also useful at intermediate SNR's.

The fact that \eqref{approx1} can be tighter than \eqref{ADT} also implies that employing an unnecessarily large number of relays with the compress-and-forward type of strategies in \cite{AvDigTse09, OD10, LKGC09} can indeed deteriorate rather than improve the communication rate. Recall that the result in \eqref{approx1} is obtained by applying these strategies with a carefully chosen subset of $k$ relays, while \eqref{ADT} is obtained by using the same strategy with all the $N$ relays. Motivated by this observation, recent work \cite{bobbie,ayan} has demonstrated the need to optimize the quantization levels in these strategies which allows to achieve the information-theoretic cutset upper bound on the capacity of the $N$-relay diamond network within $O(\log N)$ bits/s/Hz. More precisely, these works show that \eqref{ADT}, valid for any wireless network with $N$ relays, can be refined to
\begin{equation*}
 \overline{C}-\log (N+1)-\log N-1 \leq C\leq \overline{C}
\end{equation*}
for the $N$-relay diamond network. This new result can be readily used to tighten our simplification result in \eqref{eq:mainres} to
\begin{align*}
C_k&\,\geq\, \frac{k}{k+1}\, \overline{C}\,-\log(k+1)-\log k-1- G,
\end{align*}
by simply using the optimized quantization levels for the $k$-relay subnetwork. 



\section{Model}\label{sec:model}

We consider the Gaussian $N$-relay diamond network depicted in Fig.~\ref{fig:diamondN} where the source node $s$ wants to communicate to the destination node $d$ with the help of $N$ relay nodes. Let $X_s[t]$ and $X_i[t]$ denote the signals transmitted by the source node $s$ and the relay node $i\in\{1,\dots, N\}$ respectively at time instant $t\in\mathbb{N}$. Let $Y_d[t]$ and $Y_{i}[t]$ denote the signals received by the destination node $d$ and the relay node $i\in\{1,\dots, N\}$ respectively at time instant $t$. The transmitted signal $X_i[t]$ by relay $i$ is a causal function of the its corresponding received signal $Y_i[t]$.  The received signals relate to the transmitted signals as
\begin{align*}
Y_i[t]= h_{is}X_s[t]+ Z_i[t],\\
Y_d[t]=\sum_{i=1}^{N} h_{id}X_i[t]+ Z[t],
\end{align*}    
where $h_{is}$ denotes the complex channel coefficient between the source node and the relay node $i$ and $h_{id}$ denotes the complex channel coefficient between the relay node $i$ and the destination node. $Z_i[t],\,i=1,\dots, N$ and $Z[t]$ are independent and identically distributed white Gaussian random processes of power spectral density of $N_0/2$ Watts/Hz.  All nodes are subject to an average power constraint $P$ and the narrow-band system is allocated a bandwidth of $W$. 
 We assume that the channel coefficients are known at all the
 nodes. 


\section{Main Results}

The main result of this paper is summarized in the following theorems.
\begin{theorem}\label{thm:thm1}
Consider an arbitrary Gaussian $N$-relay diamond network. Let $C_k$ be the largest rate at which we can communicate from the source node to the destination using only $k$ out of the $N$ relays while the remaining $N-k$ relays are kept silent. Then 
\begin{align}
C_k&\,\geq\, \frac{k}{k+1}\, \overline{C}\label{eq:mainthm}\\
&\quad-\,1.3k-\,\frac{k}{k+1}\max\left(3\log N-\log\frac{27}{4}, 2\log N\right), 
\nonumber
\end{align}
where $ \overline{C}$ denotes the cut-set upper bound on the capacity of the $N$-relay network. Moreover, there exist configurations of the Gaussian $N$-relay diamond network such that 
\begin{equation}
C_k\leq \frac{k}{k+1} C+1.3k+\max\left(3\log k-\log\frac{27}{4}, 2\log k\right), \label{eq:mainprop}
\end{equation}
where $C$ is the capacity of the $N$-relay network.
\end{theorem}

\begin{remark}\label{rem:rem1} 
For the case $k=1$, we have the following  tighter bound,
$$
C_1\,\geq\, \frac{1}{2}\, \overline{C}-\,\frac{1}{2}\max\left(3\log N-\log\frac{27}{4}, 2\log N\right). 
$$
\end{remark}

The theorem states that in every Gaussian $N$-relay diamond network, there exists a subset of $k$ relays which alone provide approximately a fraction $k/(k+1)$ of the capacity of the whole network. On the other hand, there are also configurations, where each $k$-relay sub-network alone can at most provide this fraction of the capacity. The approximations are within the beamforming gain, which we upper bound by $\max\left(3\log N-\log\frac{27}{4}, 2\log N\right)$ for the $N$-relay diamond network uniformly over all possible channel configurations. The beamforming gain is relatively small when the capacity is large, and indeed is much smaller than this upper bound when channel gains are significantly different. On the other hand, the term $1.3k$ in the gap is not fundamental and reflects the gap between the rate achieved by the state-of-the art relaying strategies  \cite{AvDigTse09, OD10, LKGC09} and the cutset upper bound on the capacity of the diamond network with $k$ relays.\footnote{For example, using 
improved relaying strategies from recent results in \cite{bobbie,ayan}, it can be readily sharpened from $1.3k$ to $2\log k$.}

\medbreak

A key ingredient in the above results is the fact that compress-and-forward type of strategies in \cite{AvDigTse09, OD10, LKGC09} can achieve the cut-set upper bound on the capacity of any arbitrary diamond relay network within a gap that is linear in the number of relay nodes utilized, and independent of the channel configurations and the operating SNR. We next show that an amplify-and-forward strategy fails to provide such a universal performance guarantee over the channel configurations, and its performance is approximately bounded by the capacity of the best relay alone.
    
\begin{theorem}\label{thm:thm2} In any Gaussian $N$-relay diamond network, the rate $C_{AF}$ achieved by amplify-and-forward at the $N$ relays is bounded by
$$
C_{AF}\leq C_1 + 2 \log N,
$$
where $C_1$ is the capacity provided by routing over the best relay. 
\end{theorem}  

Finally, we address the algorithmic complexity of discovering a high-capacity $k$-relay subnetwork in Theorem~\ref{thm:thm1}.

\begin{theorem}\label{thm:thm3} A constant gap approximation to the capacity of the Gaussian $N$-relay diamond network can be computed in $O(N\log N)$ running time. The $k$-relay subnetwork satisfying \eqref{eq:mainthm} can be discovered in $O(kN)$ running time, given the configuration of the network and the approximation to the cutset upper bound.
\end{theorem}

Theorem~\ref{thm:thm1} is proven in \ref{sec:k}, Theorem~\ref{thm:thm2} is proven in Section~\ref{sec:AF}, and Theorem~\ref{thm:thm3} is proven in Section~\ref{sec:comp}. The following section derives a simple approximation to the cutset upper bound on the capacity of the $N$-relay diamond network, which forms the basis for all these results.
%
%

\section{Approximating the Cut-Set Upper Bound} \label{sec:cutset}

In this section we derive upper and lower bounds on the cut-set upper bound,
that essentially reduce calculating its value to a purely combinatorial problem.

Let $\N \dot= \{ 1,2,\cdots,N \}$ and  for a subset $\Lambda\subseteq \N$, $\overline{\Lambda}\dot= \N\setminus\Lambda$. By the cut-set upper bound \cite[Theorem 14.10.1]{TC91}, the capacity $C$ of the network is upper bounded by,
\begin{equation}\label{eq:cutset}
C\leq \overline{C}\dot=\max_{X_s,X_1,\dots, X_N} \min_{\Lambda\subseteq [N]} I(X_s, X_\Lambda; Y_d, Y_{\overline{\Lambda}}\,|\, X_{\overline{\Lambda}})
\end{equation}
where the maximization is over the joint probability distribution of the random variables $X_s$ and $X_1,\dots, X_N$ satisfying the power constraint $P$. For a set $S\subseteq[N]$, $X_S$ denotes the corresponding collection of random variables, i.e $ X_S\dot=\{X_i\}_{i\in S}$.

\subsection{An Upper Bound on the Cut-Set Upper Bound} \label{sec:cutsetub}
The cut-set upper bound in \eqref{eq:cutset} can be upper bounded by exchanging the order of maximization and minimization in \eqref{eq:cutset}. For each cut $\Lambda$, the resulting maximization of the mutual information can be upper bounded by the capacities of the SIMO (single input multiple output) channel between $s$ and nodes in $\overline{\Lambda}$ and the MISO (multiple input single output) channel between nodes in $\Lambda$ and $d$. We have,
\begin{align*}
\overline{C}&\leq \min_{\Lambda\subseteq [N]} \sup_{X_s,X_{\Lambda},X_{\overline{\Lambda}}} I(X_s, X_\Lambda; Y, Y_{\overline{\Lambda}}\,|\, X_{\overline{\Lambda}})
\\
&=\min_{\Lambda\subseteq [N]} \sup_{X_s} I(X_s; Y_{\overline{\Lambda}})+ \sup_{X_{\Lambda}} I(X_\Lambda; \sum_{i\in\Lambda} h_{id} X_i+Z),\\
&\leq\min_{\Lambda\subseteq [N]} C_{SIMO}(s; \overline{\Lambda})+ C_{MISO}(\Lambda; d).
\end{align*}
The capacities of the corresponding SIMO and MISO channels are well-known \cite{TV05}. Plugging these expressions yields
\begin{align}
\overline{C}&\leq \min_{\Lambda \subseteq \N}\,\, \log \Big( 1 + \SNR \sum_{i \in \overline{\Lambda}} |h_{is}|^2 \Big)\nonumber\\
&\qquad\qquad\qquad+ \log \Big( 1 + \SNR \big( \sum_{i \in \Lambda} |h_{id}| \big)^2 \Big) \label{eq:cutset3}
\end{align}
where $\SNR\dot= \frac{P}{N_0 W}$. We will further develop a simple upper bound on this expression by bounding each term in the above summations by the maximum of the terms that are summed. 
This gives us the upper bound,
\begin{equation}\label{eq:ub}
\overline{C} \leq \min_{\l \subseteq \N} \left( \max_{i \in \l}\,\, R_{id} + \max_{i \in \lbar}\,\, R_{is} \right) + G,
\end{equation}
where $R_{id}=\log \lp 1+\SNR\, |h_{id}|^2 \rp$ and $R_{is}=\log \lp 1+\SNR\,  |h_{is}|^2 \rp $ are the capacities of the corresponding point-to-point channels and\footnote{Note that the $N$-relay diamond network can be equivalently characterized in terms of these point-to-point channel capacities.}
$$
G\dot=\max \left( 3\log N - \log \frac{27}{4}, 2\log N\right).
$$
A detailed derivation of the upper bound in this section can be found in Appendix~\ref{app:app1}.

\subsection{A Lower Bound on the Cut-Set Upper Bound} \label{sec:cutsetlb}

The cut-set upper bound $\overline{C}$ above can be lower bounded by choosing $X_s, \{X_i\}_{i\in\n}$ to be independent circularly-symmetric Gaussian random variables with variance $P$, in which case  
\begin{align*}
I(X_s &, X_\Lambda; Y, Y_{\overline{\Lambda}}\,|\, X_{\overline{\Lambda}})\\
&=\log \Big( 1 + \SNR \sum_{i \in \Lambda} |h_{id}|^2 \Big)+ \log \Big( 1 + \SNR \sum_{i \in \overline{\Lambda}} |h_{is}|^2 \Big). 
\end{align*}
Retaining only the maximum terms in the summations, we obtain
\begin{equation}\label{eq:lbCG}
\overline{C}\geq \min_{\Lambda\subseteq [N]} \left( \max_{i \in \Lambda}
R_{id}+ \max_{i \in \overline{\Lambda}} R_{is} \right).
\end{equation}
Note that this lower bound for $\overline{C}$ differs from the upper bound in \eqref{eq:ub} only by the gap term $G$. This implies that within a gap of $G$ bits/s/Hz, the cut-set upper bound on the capacity of the $N$-relay diamond network behaves like the lower bound in \eqref{eq:lbCG}. Since recent results \cite{AvDigTse09,OD10,LKGC09,bobbie,ayan} show that the actual capacity of the network is within a constant gap to the cutset upper bound, this also provides an approximation to the capacity of the $N$-relay diamond network, i.e.,
\begin{equation}\label{eq:approxcap}
C\approx \min_{\Lambda\subseteq [N]} \left( \max_{i \in \Lambda}
R_{id}+ \max_{i \in \overline{\Lambda}} R_{is} \right).
\end{equation}
This reveals a peculiar combinatorial structure for the capacity  of the diamond network in terms of the point-to-point capacities of the individual channels. Our main result is based on exploiting this combinatorial structure.

\subsection{The Cut-Set Upper Bound for a $k$-Relay Sub-network}

Consider a subset $\Gamma\subseteq\N$ of the relay nodes such that $|\Gamma|=k$. Let $C_\Gamma$ be the capacity of the $k$-relay diamond sub-network where the source node $s$ wants to communicate to the destination node $d$ by using only these $k$ relay nodes.  The rest $N-k$ relays are not used. The cut-set upper bound on the capacity of this  $k$-relay network yields
\begin{equation}\label{cutsetGamma}
C_\Gamma\leq \overline{C}_\Gamma\,\dot=\sup_{X,X_\Gamma} \min_{\Lambda\subseteq \Gamma} I(X, X_\Lambda; Y, Y_{\Gamma\setminus\Lambda}\,|\, X_{\Gamma\setminus\Lambda}).
\end{equation}
Note that \eqref{eq:ub}  and \eqref{eq:lbCG} can be applied to $\Gamma$ to obtain correspondingly upper and lower bounds on $\overline{C}_\Gamma$.

Among all $\Gamma\subseteq\N$ with $|\Gamma|=k$, consider the one that has the largest cut-set upper bound $\overline{C}_\Gamma$. Let $\overline{C}_k$ denote the cut-set upper bound on the capacity of this this sub-network. Formally, we define
\begin{equation}\label{eq:barCk}
\overline{C}_k=\max_{\substack{\Gamma \subseteq \N \\ |\Gamma|=k}}\overline{C}_\Gamma.
\end{equation}
Combining \eqref{eq:lbCG} and \eqref{eq:barCk}, we have \begin{equation}\label{eq:lb}
\overline{C}_k\geq\max_{\substack{\Gamma \subseteq \N \\ |\Gamma|=k}}  \min_{\Lambda\subseteq \Gamma} \,\, \left( \max_{i \in \Lambda}
R_{id}+ \max_{i \in \Gamma\setminus\Lambda} R_{is} \right).
\end{equation}
Let $C_k$ be the capacity of the best $k$-relay sub-network. In the sequel, we will be interested in lower bounding $C_k$ in terms of $\bar{C}$, the cutset upper bound on the capacity of the network. For this, we will first relate $C_k$ to $\overline{C}_k$ and then make use of the above lower bound for $\overline{C}_k$.

\section{$k$ Relays Approximately Achieve $\frac{k}{k+1}$ Fraction of the Capacity}\label{sec:k}

In this section, we prove Theorem~\ref{thm:thm1}. However, before going into the formal proof, let us illustrate the main idea for $k=1$. Assume the capacity $C$ of the $N$-relay diamond network were given exactly by \eqref{eq:approxcap}, while the capacity obtained by using relay $i\in[N]$ alone is given by
$$
C_i=\min(R_{is}, R_{id}). 
$$
Note that this is the capacity approximation in \eqref{eq:approxcap} evaluated for a single relay ($N=1$), but in this particular case it indeed corresponds to the exact capacity of a single relay (2-hop) network. Can we argue that there exists a relay $i\in[N]$ such that  $C_i\geq C/2$? This is  easy. If this were not the case, it would imply that
$$
\forall i\in[N],\qquad\textrm{either}\qquad R_{is}< \frac{C}{2}\quad\textrm{or}\quad R_{id}< \frac{C}{2}.  
$$
This would allow us to construct a cut of the network $\Lambda$ which crosses only  links with capacities strictly smaller than $C/2$, both on the source side and the destination side, i.e., $R_{id}<C/2$ $\forall i\in \Lambda$ and $R_{is}<C/2$ $\forall i\in \overline{\Lambda}$. Hence the value of this cut is strictly smaller than $C$ and this contradicts with our initial assumption that the capacity of the $N$-relay diamond network is $C$. Therefore, there exists at least one relay $i\in[N]$ such that  $C_i\geq C/2$. To prove the converse statement in Theorem~\ref{thm:thm1},  we need to create examples where each relay alone only provides half the capacity of whole network: consider a configuration where $R_{is}=C/2$ and $R_{id}=C$ for some of the relays and $R_{is}=C$ and $R_{id}=C/2$ for the rest. The capacity of the whole network is $C$ by \eqref{eq:approxcap}, while each relay alone can  only provide capacity $C/2$. 

The formal proof of Theorem~\ref{thm:thm1} is based on the following two technical lemmas.

\begin{lemma}\label{lem:lem1}
Let $R_{id}$ and $R_{is}$ be arbitrary positive real numbers for $i=1,2,\cdots,N$. For $k\in \N$, let
\begin{equation}\label{eq:defrk}
r_k\dot= \frac{\displaystyle\max_{\substack{\Gamma \subseteq \N \\ |\Gamma|=k}} \min_{\l \subseteq \Gamma} \lp \max_{i \in \l} R_{id} + \max_{i \in \Gamma\setminus\Lambda} R_{is} \rp}{\displaystyle \min_{\l \subseteq \N} \lp \max_{i \in \l} R_{id} + \max_{i \in \lbar} R_{is} \rp }.        
\end{equation}
Then,
$$
r_k\geq\frac{k}{k+1}.
$$
\end{lemma}

\begin{figure}[t]
\begin{center}
\includegraphics[scale=0.7]{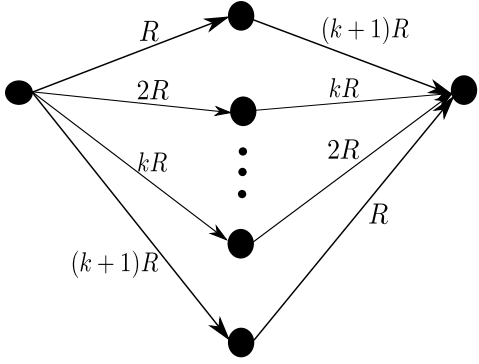}
\caption{A $(k+1)$-relay diamond network where every subset of $k$ relays achieve approximately $\frac{k}{k+1}$ of the capacity. The labels indicate the capacity of the corresponding links.}
\label{fig:MHlbk}
\end{center}
\end{figure}

\medbreak
\begin{lemma}\label{lem:lem2}
Let $R_{is} = i\,R$ and $R_{id} = (k+2-i)\,R$ for $i \in [k+1]$ where $R$ is an arbitrary positive number. Let $r_k$ be defined as in \eqref{eq:defrk} with $N=k+1$. Then,
$$
r_k=\frac{k}{k+1}.
$$
\end{lemma}
The configuration in Lemma~\ref{lem:lem2} is depicted in Fig.~\ref{fig:MHlbk}.

\medbreak

\textit{Proof of Theorem~\ref{thm:thm1}:} 
From \eqref{eq:ub} and \eqref{eq:lb}, we have
$$
\frac{\overline{C}_k}{\overline{C}-G}\geq r_k.
$$
Combining this with the result of Lemma~\ref{lem:lem1}, we obtain
\begin{equation}\label{eq:eq0}
\overline{C}_k\geq \frac{k}{k+1} \overline{C} - \frac{k}{k+1} G.
\end{equation}
This proves that in every $N$ relay diamond network, there exists a subset of $k$ relays, such that the cut-set upper bound on the capacity of the corresponding $k$ relay subnetwork is lower bounded by approximately a fraction $\frac{k}{k+1}$ of the cut-set upper bound on the capacity of the whole network. Let $\Gamma^*\subset[N]$ be the maximizing term in \eqref{eq:barCk}, i.e., $ \overline{C}_{\Gamma^*}= \overline{C}_k$, and let $C_{\Gamma^*}$ be the actual capacity of this network. From \cite{LKGC09},
$C_{\Gamma^*} \geq \overline{C}_{\Gamma^*} - 1.3k,$ for any $k$-relay network, which is achieved by a noisy network coding strategy generalizing the quantize-map-and-forward strategy of \cite{AvDigTse09}. Let $C_k$ be the capacity of the best $k$-relay subnetwork. Since $C_k\geq C_{\Gamma^*}$ by definition, we have
$$
C_k \geq \overline{C}_k - 1.3k.
$$
Together with \eqref{eq:eq0} this yields the result \eqref{eq:mainthm} in Theorem~\ref{thm:thm1}.

\smallbreak
Next, we prove the existence of a $(k+1)$-relay diamond network where the capacity of each $k$-relay sub-network satisfies \eqref{eq:mainprop}, i.e., for now we assume $N=k+1$. To prove this, we require an upper bound on $C_k$ and a lower bound on $C$. The lower bound on $C$ can be obtained by  combining \eqref{eq:lbCG} with the fact that $C\geq \overline{C}-1.3(k+1)$ from \cite{LKGC09} (since $N=k+1$), which yields 
\begin{equation}\label{eq:Clb}
C\geq \min_{\Lambda\subseteq \N} \left( \max_{i \in \Lambda}
R_{id}+ \max_{i \in \overline{\Lambda}} R_{is} \right)\,-1.3(k+1).
\end{equation}
On the other hand, applying \eqref{eq:ub} for any $\Gamma\subseteq[k+1]$ s.t $|\Gamma|=k$, we obtain
$$
\overline{C}_\Gamma \leq \min_{\l \subseteq \Gamma} \,\, \left( \max_{i \in \l} R_{id} + \max_{i \in \lbar} R_{is} \right) + G_k,
$$
where 
$
G_k\dot=\max \left( 3\log k - \log \frac{27}{4}, 2\log k\right).
$
Therefore, 
\begin{equation}\label{eq:Ckub}
\overline{C}_k \leq \max_{\substack{\Gamma \subseteq [k+1] \\ |\Gamma|=k}}  \min_{\l \subseteq \Gamma} \left( \max_{i \in \l} R_{id} + \max_{i \in \lbar} R_{is} \right)  + G_k.
\end{equation}
Combining \eqref{eq:Clb} and \eqref{eq:Ckub}, we obtain
$$
\frac{\overline{C}_k-G_k}{C+1.3(k+1)}\leq r_k.
$$
Lemma~\ref{lem:lem2} demonstrates a configuration where $r_k=\frac{k}{k+1}$. For such configurations, the above inequality yields
$$
\overline{C}_k\leq \frac{k}{k+1}C+1.3k+G_k.
$$
Since $C_k\leq \overline{C}_k$, this proves that there exist $k+1$-relay diamond networks such that the capacity of each $k$-relay subnetwork satisfies the bound \eqref{eq:mainprop} in Theorem~\ref{thm:thm1}. However, Theorem~\ref{thm:thm1} claims the existence of $N$-relay diamond networks where each $k$-relay subnetwork satisfies \eqref{eq:mainprop}. To extend the proof to any $N>k$, simply consider augmenting the $k+1$ relay diamond network of Fig.~\ref{fig:MHlbk} by adding relay nodes with zero capacities. Whatever holds for the $k+1$-relay network also holds for this trivially augmented $N$-relay network.   This completes the proof of Theorem~\ref{thm:thm1}. $\hfill\square$
\bigbreak

We will next prove Lemma~\ref{lem:lem1} for the case $k=1$ and $k=2$. The proof of Lemma~\ref{lem:lem1} for $k>2$ and the proof of Lemma~\ref{lem:lem2} are provided in Appendix~\ref{app:app2}. 
\bigbreak
\textit{Proof of Lemma~\ref{lem:lem1}:} We introduce the following notation. Let
\begin{align}
\w(\g) &\dot= \min_{\l \subseteq \Gamma} \left( \max_{i \in \l} R_{id} + \max_{i \in \lbar} R_{is} \right)\\
 \wbar &\dot= \min_{\l \subseteq \N} \left( \max_{i \in \l} R_{id} + \max_{i \in \lbar} R_{is} \right),\label{eq:wbar}
\end{align}
and
$
 \w_k \dot=  \max_{\substack{\Gamma \subseteq \N \\ |\Gamma|=k}} \w(\g). 
$
Note that $r_k$ in Lemma~\ref{lem:lem1} is defined as $r_k=\frac{w_k}{\wbar}$. 

The first thing we note is that $r_k \leq 1$. This follows from the fact that every subset of $\Gamma$ is necessarily a subset of $\N$,i.e., if $\l \subseteq \Gamma$ then $\l \subseteq \N$ and $\Gamma\setminus\l\,\subseteq\,\N\setminus\l$. Therefore, the value of each cut $\l\subseteq \Gamma$ in $\Gamma$ is smaller than or equal to the value of the same cut in $\N$. The same reasoning also implies that for $k_1 \geq k_2$ we have $r_{k_1} \geq r_{k_2}$. Both properties are to be naturally satisfied by a capacity function: by  using more relays we can only increase the capacity.

\begin{list}{\labelitemi}{\leftmargin=1em}
\item 
For $k=1$, the lemma claims that $w_1\geq\frac{1}{2}{\wbar}$. Since
$$
w_1=\max_{i \in \N}  \min \left(R_{id}, R_{is}\right),
$$
this is equivalent to saying that $\exists\, i\in[N]$ s.t. $R_{id}\geq\frac{1}{2}{\wbar}$ and $R_{is}\geq\frac{1}{2}{\wbar}$. We will prove this by contradiction. Assume
\begin{equation}\label{eq:ass}
\forall i\in\N,\,\, R_{id}<\frac{1}{2}{\wbar}\quad\textrm{or}\quad R_{is}<\frac{1}{2}{\wbar}.
\end{equation}
Let $\l_0 = \left\{ i \in \N : R_{id} < \frac{1}{2}{\wbar}\right\}$. The assumption in \eqref{eq:ass} implies that $R_{is}<\frac{1}{2}{\wbar},\,\,\forall i\in \overline{\l}_0$. Note that $\wbar$ in \eqref{eq:wbar} can be upper bounded by considering only the cut $\Lambda_0$ among all possible cuts $\Lambda\subseteq\N$. We obtain
$$
\wbar\leq \max_{i \in \l_0} R_{id} + \max_{i \in \lbar_0} R_{is}<\wbar 
$$
since each of the two terms are strictly smaller than $\frac{1}{2}{\wbar}$. This contradiction proves the lemma for $k=1$.
\smallbreak
\item 
For $k=2$, the lemma claims that $w_2\geq\frac{2}{3}{\wbar}$. We can prove this by establishing a number of properties for a network with $\wbar$.
\smallbreak
\emph{Property:} $\exists\, p\in[N]$ s.t. $R_{ps}\geq\frac{2}{3}{\wbar}$ and $R_{pd}\geq\frac{1}{3}{\wbar}$. 
\smallbreak
We prove this by contradiction. Assume $$
\forall i\in\N,\,\, R_{is}<\frac{2}{3}{\wbar}\quad\textrm{or}\quad R_{id}<\frac{1}{3}{\wbar}.
$$
Consider the cut $\l_0 = \left\{ i \in \N : R_{id} < \frac{1}{3}{\wbar}\right\}$. Then $R_{is}<\frac{2}{3}{\wbar},\,\,\forall i\in \overline{\l}_0$. Considering only the cut $\Lambda_0$ we obtain
$$
\wbar\leq \max_{i \in \l_0} R_{id} + \max_{i \in \lbar_0} R_{is}<\wbar, 
$$
which is a contradiction.

We next proceed by investigating two separate cases:
\begin{itemize}
\item \emph{Case 1:} $R_{pd}\geq \frac{2}{3}{\wbar}$. Then, the proof of the lemma is complete since we have $w_2\geq w_1\geq \frac{2}{3}{\wbar}$.
\smallbreak
\item \emph{Case 2:} $R_{pd}< \frac{2}{3}{\wbar}$. Then we establish the following property:
\smallbreak
\emph{Property:} $\exists \, m\in[N],\  m \neq p$ s.t. $R_{ms}\geq\frac{1}{3}{\wbar}$ and $R_{md}\geq\frac{2}{3}{\wbar}$. 
\smallbreak
Again, we can prove this property by contradiction. Assume the contrary and consider $\l_1 = \left\{ i \in \N : R_{id} < \frac{2}{3}{\wbar}\right\}$. Note that $p\in\l_1$ since we are in Case~$2$ and $R_{is}<\frac{1}{3}{\wbar},\,\,\forall i\in \overline{\l}_1$. The value of the cut $\l_1$ is strictly smaller than $\wbar$, which is a contradiction.
\smallbreak

Finally, consider the $2$-relay sub-network composed of $m$ and $p$. It can be easily verified that $\w(\{m,p\})\geq \frac{2}{3}{\wbar}$, completing the proof of the lemma for $k=2$. 
\end{itemize}

\end{list}

The proof of the lemma for the general case follows similar lines. The main idea is to show that given any arbitrary real numbers $R_{id}$ and $R_{is}$ for $i=1,2,\cdots,N$, we can gradually discover a $k$-relay subnetwork $\Gamma^*$ such that $C_{\Gamma^*}\geq\frac{k}{k+1}\wbar$. \hfill$\square$

\section{Algorithmic Complexity} \label{sec:comp}

Given an arbitrary $N$-relay diamond network, characterized by the point-to-point capacities of the individual links $R_{is}, R_{id}$, $i\in\N$, can we efficiently discover a $k$-relay subnetwork whose capacity satisfies \eqref{eq:mainthm}? In this section, we prove Theorem~\ref{thm:thm3}.

Note that from the proof of Theorem~\ref{thm:thm1}, the $k$-relay subnetwork $\Gamma^*\subseteq \N$ whose capacity $C_{\Gamma^*}$ satisfies \eqref{eq:mainthm} is the one for which $w(\Gamma^*)/\wbar\geq\frac{k}{k+1}$, where $w(\Gamma^*)$ and $\wbar$ are defined in \eqref{eq:wbar}. The proof of Lemma~\ref{lem:lem1} suggests a natural algorithm to discover this network.

\begin{itemize}
\item For $k=1$, the lemma proves that
$$ 
\exists i\in\N,\,\, R_{id}\geq\frac{1}{2}{\wbar}\quad\textrm{and}\quad R_{is}\geq\frac{1}{2}{\wbar}.
$$
This node $i$ can be discovered by making $2N$ comparisons in the worst case.
\item For $k=2$, the lemma first proves that
$$ 
\exists p\in\N,\,\, R_{pd}\geq\frac{2}{3}{\wbar}\quad\textrm{and}\quad R_{ps}\geq\frac{1}{3}{\wbar}.
$$
Then either $R_{ps}\geq\frac{2}{3}{\wbar}$ or 
$$ 
\exists m\in\N,\,m\neq p\quad\textrm{and}\quad R_{md}\geq\frac{1}{3}{\wbar}\quad\textrm{and}\quad R_{ms}\geq\frac{2}{3}{\wbar}.
$$
We can follow this flow to discover relays $p$ and $m$ for which we have $\w(\{m,p\})\geq \frac{2}{3}{\wbar}$. $p$ can be discovered in at most $2N$ comparisons. An extra comparison determines whether $R_{ps}\geq\frac{2}{3}{\wbar}$ or $\frac{1}{3}{\wbar}\leq R_{ps}<\frac{2}{3}{\wbar}$. In the first case, the algorithm terminates. Otherwise, we need at most $2(N-1)$ additional comparisons to discover $m$. This yields $4N-1$ comparisons in the worst case.

\item For $2<k<N$, the proof of the lemma in Appendix~\ref{app:app2} shows that any positive real numbers $R_{is}$, $R_{id}$,  $i \in \N$ can be either arranged as 
\begin{itemize}
\item $R_{Ns}\geq \frac{k}{k+1} \wbar \quad\textrm{and}\quad R_{Nd}\geq \frac{k}{k+1} \wbar,
$
\end{itemize}
or
\begin{itemize}
\item $R_{Ns} \geq \frac{k}{k+1} \wbar$ and $\frac{k-a+1}{k+1} \wbar > R_{Nd} \geq \frac{k-a}{k+1} \wbar$ for some  $a \in \mathbb{N}$ such that $1 \leq a \leq k-1$,
\item and for $1 \leq r \leq l$, $\frac{a_r+1}{k+1}\wbar > R_{rs} \geq \frac{a_r}{k+1} \wbar$ and $R_{rd} \geq \frac{k-a_{r-1}}{k+1} \wbar$ for some  $l\in \mathbb{N}$ such that $1\leq l\leq k-2$, and $a_0, a_1, \dots, a_{l}\in \mathbb{N}$ such that  $a_0=0<a_1<\dots<a_{l-1}<a_{l}<a$,
\item and $R_{l+1,d} \geq \frac{a}{k+1} \wbar$ and $R_{yd} \geq \frac{k-a_{l}}{k+1} \wbar$.
\end{itemize}
For these $l+2\leq k$ nodes $\Gamma^*=[l+1]\cup \{N\}$, we have $w(\Gamma^*)\geq \frac{k}{k+1}\wbar$.

\medbreak
The flow in the proof of the lemma suggests a natural algorithm to make this arrangement. 

\begin{itemize}
\item[(a)] Find the node $i\in \n$ such that 
$$R_{Ns}\geq \frac{k}{k+1} \wbar \quad\textrm{and}\quad R_{Nd}\geq \frac{1}{k+1} \wbar,$$
and label it node $N$.

\item[(b)] Determine $a$ such that $1\leq a\leq k-1$ and 
$$R_{Ns}\geq \frac{k}{k+1} \wbar \,\,\textrm{and}\,\, \frac{k-a+1}{k+1} \wbar >R_{Nd}\geq \frac{k-a}{k+1} \wbar.$$
\item[(c)] If $a=1$, terminate the algorithm and declare $\Gamma^*=\{N\}$. 
Otherwise, set $a_0=0$.
\item[(d)] For $1\leq r\leq k-2$,
\begin{itemize}
\item[(d-1)] Find the node $i\in [r,N-1]$ such that $$R_{is}\geq \frac{a_{r-1}+1}{k+1} \wbar \,\,\textrm{and}\,\, R_{id}\geq \frac{a_{r-1}}{k+1} \wbar,
$$
and label it node $r$. 
\item[(d-2)] Determine $a_r$ such that $a_{r-1} < a_{r}\leq a$, and $$\hspace*{-0.2cm}\frac{a_{r}+1}{k+1} \wbar> R_{rs}\geq \frac{a_{r}}{k+1} \wbar \,\,\textrm{and}\,\, R_{rd}\geq \frac{k-a_{r-1}}{k+1} \wbar.$$ 

\item[(d-3)] If $a_{r}=a$, terminate the algorithm and declare $\Gamma^*=[r]\cup\{N\}$. Otherwise set $r\leftarrow r+1$.
\end{itemize}
\end{itemize}
 
The total number of comparisons to be made by the algorithm can be upper bounded as follows:

\begin{itemize}
\item Step (a): at most $2N$ comparisons
\item Step (b): at most $k-1$ comparisons
\item Step (d-1): at most $2(N-r)$ comparisons
\item Step (d-1): at most $k-1-r$ comparisons
\end{itemize}
Assuming that step (d) makes the maximum number of iterations $k-1$, the total number of comparisons to be made by the algorithm is upper bounded by 
$$
2N+ (k-1)+\sum_{r=1}^{k-1} 2(N-r)+(k-1-r)=2Nk-\frac{(k-1)k}{2}.
$$ 
\end{itemize}

However, the  above discussion assumes that $\wbar$ is given. Given the set of real numbers $R_{is}, R_{id}$, $i=1,\dots, N$, a straightforward approach to computing $\wbar$ in \eqref{eq:wbar} requires the evaluation of $2^N$ cuts, while computing the value of each cut requires $N$ comparisons. Instead, the following algorithm allows to compute $\wbar$ in $N\log N$ running time. 

\begin{figure}[t]
\begin{center}
\includegraphics[scale=0.5]{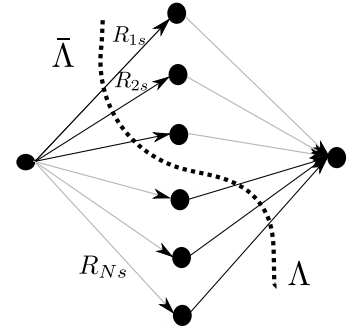}
\caption{The minimum cut on a configuration such that $R_{1s}\leq\dots\leq R_{Ns}$.}
\label{fig:wbar}
\end{center}
\end{figure}

First, sort (rearrange) the nodes in the order of increasing $R_{is}$, i.e., $R_{1s}\leq\dots\leq R_{Ns}$. For this sorted configuration, observe that the cut with the minimum value in \eqref{eq:wbar}, i.e., the cut $\Lambda^*$  for which
$$
 \wbar = \max_{i \in \l^*} R_{id} + \max_{i \in \lbar^*} R_{is} ,
$$
is necessarily of the form in Figure~\ref{fig:wbar}. More precisely,
$
\l^*=\l_m\triangleq [m+1,N]
$
and $\lbar^*=[m]$ for some $1\leq m\leq N$. This is easy to see: consider any cut $\l\subseteq[N]$ not necessarily of the form in Figure~\ref{fig:wbar}. Let $m$ be the node in $\lbar$ with the largest index, i.e., $m=\max\{i\in\lbar\}$ and let $
\l_m=\{m+1,\dots,N\}$. We have
$$
 \max_{i \in \l_m} R_{id} + \max_{i \in \lbar_m} R_{is}\leq \max_{i \in \l} R_{id} + \max_{i \in \lbar} R_{is}.
$$
The second terms are equal because $R_{is}$ are sorted in increasing order and the first term can be only smaller for $\l_m$ since it is a subset of $\l$ by construction. This reduces the number of candidate cuts for the min cut from $2^N$ to $N$. 

In other words, the mincut can be calculated by making $N$ comparisons of two numbers:  the maximum value $R_{is}$,  $i\in \lbar_m$, with the maximum $R_{id}$, with $i\in \l_m$, for  $\l_m=[m+1,N]$, $m=0,\ldots,N$. 
Assume that the $R_{is}$ values are sorted as previously described - this can be  done using $N\log N$  comparisons, for example with the heap sort algorithm.
Thus for the  set $\lbar_m$, the value we would use is $R_{ms}$. 
But we can also keep a sorted heap of the $R_{id}$ values, that again can be created using $N\log N$ operations. Then for $\l_1$ we would use the max value, for $\l_2$ the max value after removing $R_{1d}$, etc. That is, we can take advantage of the fact that each subset of $R_{id}$'s would also be ordered, to extract the max value of the subset. Thus in total of $N+2\log N$
comparisons, we can compute $\omega$.


 This implies that with at most $(2k+1)N+2N\log N$ comparisons we can compute $\wbar$, a constant gap approximation to the capacity of the $N$-relay diamond network, and identify a $k$-relay subnetwork that  approximately achieves a fraction $k/(k+1)$ of $\wbar$. This completes the proof of Theorem~\ref{thm:thm3}.

\section{Amplify-and-Forward with $N$ Relays vs. Routing over the Best Relay}  \label{sec:AF}
In this section,  we derive an upper bound on the rate achieved by amplify-forward over the Gaussian N-relay diamond network in terms of the capacity of the best relay. With amplify-forward, the transmitted signals from the relay nodes are nothing but the scaled versions of the received signals from the source, $X_i[t] = \beta_i Y_i[t]$. This induces a  point-to-point link between the source node and destination given by,
\begin{equation*}
Y_d[t]=\left( \sum_{i=1}^{N} h_{id}h_{is} \beta_i \right) X_s[t] + \left(Z[t]+  \sum_{i=1}^{N} h_{id} \beta_i Z_i[t] \right).
\end{equation*}
Using the familiar capacity expression for a point-to-point AWGN channel, we get
\begin{align}
C_{AF} = \log \left( 1+ \frac{\left|\sum_{i=1}^{N} h_{id}h_{is} \beta_i \right|^2\SNR}{			1+ \sum_{i=1}^{N} |h_{id}|^2 |\beta_i|^2 } \right).\label{eq:rateAF}
\end{align}
The $\beta_i$'s in the above expression can be optimized to get the largest communication rate subject to the power constraint at the relays. Since $\EE[|X_i|^2]\leq P$, we can write
\begin{equation*}
|\beta_i|^2 =  \frac{\SNR}{1+|h_{is}|^2\SNR} |\alpha_i|^2,
\end{equation*}
where $|\alpha_i| \leq 1$ for each $i$. Next, we first upper bound the rate in \eqref{eq:rateAF} and then express it in terms of the new variables $\alpha_i$. Applying the Cauchy-Schwarz inequality on the numerator of the fractional term inside the logarithm, we get
\begin{align*}
C_{AF} &\leq \log \left( 1 + \frac{N \sum_{i=1}^{N} |h_{id}|^2 |h_{is}|^2|\beta_i|^2\,\SNR} { 1 + \sum_{i=1}^{N} |h_{id}|^2 |\beta_i|^2 }  \right) \\
&\leq \log \left( 1 + \frac{ N^2 \max_{i\in \N} |h_{id}|^2 |h_{is}|^2|\beta_i|^2\,\SNR }{ \max \left( 1, \max_{i\in \N} |h_{id}|^2 |\beta_i|^2 \right)}  \right).
\end{align*}
The second inequality is obtained by upper bounding each term of the sum in the numerator by the maximum term and taking only the maximum element for the sum in the denominator. In terms of $\alpha_i$, this last upper bound can be expressed as 
$$
C_{AF}\leq \log \left( 1 + \frac{ N^2 \max_{i\in \N}  \frac{|h_{id}|^2 |h_{is}|^2|\alpha_i|^2\SNR^2}{1+|h_{is}|^2\SNR}  }{ \max \left( 1, \max_{i\in \N}   \frac{|h_{id}|^2|\alpha_i|^2\SNR}{1+|h_{is}|^2\SNR}  \right)}  \right).
$$
In Lemma~\ref{lem:lem_a_f} below, we show that for any arbitrary positive real numbers  $u_{id}$, $u_{is}$ and $0\leq b_i\leq 1$,  $i=1,2,\cdots,N$, we have
\begin{equation}\label{eq:lem_a_f}
\max \left( 1, \max_{i \in \N} \frac{u_{id} b_i}{1+u_{is}} \right) \max_{i \in \N} \left(\min(u_{id},u_{is}) \right) \geq \max_{i\in \N} \frac{b_i u_{id} u_{is}}{1+u_{is}}
\end{equation}
Plugging $u_{id}=|h_{id}|^2\SNR$, $u_{is}=|h_{is}|^2\SNR$ and $b_i=|\alpha_i|^2$ in this relation, we get
\begin{align*}
C_{AF} &\leq \log \left( 1 + N^2 \max_{i \in \N} \min(|h_{id}|^2\SNR,|h_{is}|^2\SNR)  \right) \\
				&\leq \max_{i \in \N}\min(R_{is}, R_{id}) +2 \log N \\
				&=C_1 + 2 \log N.
\end{align*}

This proves Theorem~\ref{thm:thm2}. Lastly, we prove the inequality in \eqref{eq:lem_a_f}.

\begin{lemma} \label{lem:lem_a_f}
Let $u_{id}$, $u_{is}$ be arbitrary positive real numbers and $b_i$ be a real number in the interval $\left[0, 1 \right]$ for $i=1,2,\cdots,N$. Then,
\begin{equation}\label{eq:lem3}
\max \left( 1, \max_{i \in \N} \frac{u_{id} b_i}{1+u_{is}} \right) \max_{i \in \N} \left(\min(u_{id},u_{is}) \right) \geq \max_{i\in \N} \frac{b_i u_{id} u_{is}}{1+u_{is}}.
\end{equation}
\end{lemma}
\textit{Proof of Lemma~\ref{lem:lem_a_f}:}
The expression on the left-hand side of \eqref{eq:lem3} can be rewritten as
\begin{align*}
\gamma = \max_{i \in \N} \max \{ &\min(u_{id},u_{is}), \frac{u_{id} b_i}{1+u_{is}} \min(u_{id},u_{is}), \\
																			& \min(u_{id},u_{is}) \max_{j \in \N, j\neq i} \frac{u_{jd} b_j}{1+u_{js}}  \}. 
\end{align*}
If $u_{is} < u_{id}$, $\frac{u_{id} b_i}{1+u_{is}}\min(u_{id},u_{is})=\frac{u_{id} u_{is} b_i}{1+u_{is}}$ is among the terms to be maximized in $\gamma$. If $u_{is} \geq u_{id}$, $\min(u_{id},u_{is})=u_{id}$ is among the terms to be maximized in $\gamma$ and it satisfies $u_{id} > \frac{u_{id} u_{is} b_i}{1+u_{is}}$. Therefore, we can immediately conclude that
\begin{equation*}
\gamma\geq \max_{i\in \N} \frac{b_i u_{id} u_{is}}{1+u_{is}}.
\end{equation*}

\section{Conclusions}
We showed that in an $N$-relay diamond network we can use $k$ of the $N$ relays and approximately maintain a $\frac{k}{k+1}$ fraction of the total capacity. In particular, we can use a single relay and approximately achieve half the capacity. Our proof was based on reducing the network simplification to a combinatorial problem.

\appendices
\section{An Upper Bound on the Cut-set Upper Bound\\
\emph{(detailed derivation of Section~\ref{sec:cutsetub})}}\label{app:app1}
The cut-set upper bound in \eqref{eq:cutset} can be further upper bounded by 
\begin{align}
\overline{C}&\leq \min_{\Lambda\subseteq [N]} \sup_{X_s,X_{\Lambda},X_{\overline{\Lambda}}} I(X_s, X_\Lambda; Y_d, Y_{\overline{\Lambda}}\,|\, X_{\overline{\Lambda}})\label{app:cutset0}\\
&\leq\min_{\Lambda\subseteq [N]} \sup_{X_s,X_{\Lambda}} I(X_s, X_\Lambda; \sum_{i\in\Lambda} h_{id} X_i+Z, Y_{\overline{\Lambda}})\label{app:cutset1}\\
&\leq\min_{\Lambda\subseteq [N]} \sup_{X} I(X_s; Y_{\overline{\Lambda}})+ \sup_{X_{\Lambda}} I(X_\Lambda; \sum_{i\in\Lambda} h_{id} X_i+Z),\label{app:cutset2}
\end{align}
where \eqref{app:cutset0} follows by changing the order of maximization and minimization in \eqref{eq:cutset}; \eqref{app:cutset1} follows because 
\begin{align*}
&I(X_s, X_\Lambda; Y_d, Y_{\overline{\Lambda}}\,|\, X_{\overline{\Lambda}})=I(X_s, X_\Lambda; Y_d-\sum_{i\in\overline{\Lambda}} h_{id} X_i, Y_{\overline{\Lambda}}\,|\, X_{\overline{\Lambda}})\\
&\,=h(Y_d-\sum_{i\in\overline{\Lambda}} h_{id} X_i, Y_{\overline{\Lambda}}\,|\, X_{\overline{\Lambda}})\\
&\qquad\quad-h(Y_d-\sum_{i\in\overline{\Lambda}} h_{id} X_i, Y_{\overline{\Lambda}}\,|\,X_s, X_\Lambda, X_{\overline{\Lambda}})\\
&\,=h(Y_d-\sum_{i\in\overline{\Lambda}} h_{id} X_i, Y_{\overline{\Lambda}}\,|\, X_{\overline{\Lambda}})-h(Z, Z_{\overline{\Lambda}})\\
&\leq h(Y_d-\sum_{i\in\overline{\Lambda}} h_{id} X_i, Y_{\overline{\Lambda}})-h(Z, Z_{\overline{\Lambda}})\\
&=I(X_s, X_\Lambda; \sum_{i\in\Lambda} h_{id} X_i+Z, Y_{\overline{\Lambda}}).
\end{align*}
Note that this last expression maximized over all random variables $X_s, X_{\Lambda}$ is the capacity of the point to point channel between $\{s,\Lambda\}$ and $\{\overline{\Lambda},d\}$. The capacity of this channel can be further upper bounded by the sum of the capacities of the SIMO channel between $s$ and $\{\overline{\Lambda}\}$ and the MISO channel between $\{\Lambda\}$ and $d$ which is the result stated in \eqref{app:cutset2}. Formally, this follows because
\begin{align*}
&I(X_s, X_\Lambda; \sum_{i\in\Lambda} h_{id} X_i+Z, Y_{\overline{\Lambda}})\\
&\,\,\leq h(\sum_{i\in\Lambda} h_{id} X_i+Z)+ h( Y_{\overline{\Lambda}})-h(Z)- h(Z_{\overline{\Lambda}})\\
&\,\,=I(X_s; Y_{\overline{\Lambda}})+I(X_\Lambda; \sum_{i\in\Lambda} h_{id} X_i+Z).
\end{align*}
The solutions to the maximization of these mutual informations over the imput distributions are well-know  and yield the capacities of the corresponding SIMO and MISO channels \cite{TV05}. Therefore, \eqref{app:cutset2} can be further upper bounded as
\begin{align}
\overline{C}&\leq \min_{\Lambda \subseteq \N} \Bigg(\log \Big( 1 + \SNR \sum_{i \in \overline{\Lambda}} |h_{is}|^2 \Big)\nonumber\\
&\qquad\qquad\qquad+ \log \Big( 1 + \SNR \big( \sum_{i \in \Lambda} |h_{id}| \big)^2 \Big) \Bigg)\label{app:cutset3}
\end{align}
where $\SNR\dot= \frac{P}{N_0 W}$. We will further develop a trivial upper bound on this expression. For simplicity of notation, let us introduce  $t_{is}\dot=\sqrt{\SNR}|h_{is}|$ and $t_{id}\dot=\sqrt{\SNR}|h_{id}|$. Separating the cases $\l=\emptyset$ and $\l=\N$, which correspond to the pure SIMO and pure MISO cuts respectively in \eqref{app:cutset3}, we have,
\begin{align*}
&\overline{C}\leq  \min \Bigg\{ \log \Big( 1 + \sum_{i \in \N} t_{is}^2 \Big), \log \Big( 1 + \big( \sum_{i \in \N} t_{id} \big)^2 \Big), \\
&\, \min_{\substack{\Lambda \subseteq \N \\ |\l|\neq0,N}} \Big(\log \Big( 1 + \sum_{i \in \overline{\Lambda}} t_{is}^2 \Big)+ \log \Big( 1 + \big( \sum_{i \in \Lambda} t_{id} \big)^2 \Big) \Big)\Bigg\}.
\end{align*}
Note that the variables $t_{is}$ and $t_{id}$ are real and positive. The sums over the variables $t_{id}$ and $t_{is}$ can be increased by setting each summand to the maximum of the variables that are summed. For example, using
also the fact that $\log$ is strictly increasing we can write,
\begin{equation*}
 \log \Big(  1 + \big( \sum_{i \in \l} t_{id} \big)^2 \Big)  \leq \log \big(  |\l|^2 + |\l|^2 \max_{i \in \l} t_{id}^2 \big) \text{ if }|\l|>0.
\end{equation*}
Using similar arguments we get the following inequality,
 \begin{align*}
\overline{C}\leq \min \Bigg\{ &\log \Big( 1 + \max_{i \in \N} t_{is}^2 \Big) + \log N,\\
&\log \Big( 1 + \max_{i \in \N} t_{id}^2 \Big)+ 2\log N,\\ &\min_{\substack{\Lambda \subseteq \N \\ |\l|\neq0,N}} \Big(\log \Big( 1 + \max_{i \in \overline{\Lambda}} t_{is}^2 \Big)\\
&\qquad+ \log \Big( 1 + \max_{i \in \Lambda} t_{id}^2 \Big)  + \log\big(|\l|^2 |\lbar|\big)\Big)\Bigg\}.
 \end{align*}
Let us first focus on the $\log \lp |\l|^2 |\lbar| \rp $ term. We have $|\l|+ |\lbar|=N$ and hence $$\log \lp |\l|^2 |\lbar| \rp =\log \lp N |\l|^2 -|\l|^3 \rp.$$
This term is maximized when $|\l|=\frac{2N}{3}$. Hence,
$$\log \lp |\l|^2 |\lbar| \rp \leq 3 \log N - \log \frac{27}{4}.$$
Noting that $$\log \big(  1+ \max_{i \in \Lambda} t_{id}^2 \big)= \max_{i \in \Lambda} \log \lp 1+ t_{id}^2 \rp, $$
we obtain the following upper bound,
\begin{equation}\label{app:ub}
\overline{C} \leq \min_{\l \subseteq \N} \,\, \max_{i \in \l} \log \lp 1+ t_{id}^2 \rp + \max_{i \in \lbar} \log \lp 1+ t_{is}^2 \rp  + G,
\end{equation}
where 
$$
G\dot=\max\left(3\log N - \log \frac{27}{4}, 2\log N\right).
$$

\section{A combinatorial Problem\\
\emph{(proofs of  Lemmas~\ref{lem:lem1} and \ref{lem:lem2})}
}\label{app:app2}

In addition to $\w(\Gamma)$, $\wbar$, $\w_k$ defined in Section~\ref{sec:k}, in the due analysis we also use the notation $\set{a,a+b} = \{a, a+1, \cdots, a+b\}$ for $a\geq1$ and $b\geq0$. 

\bigskip
\textit{Proof of Lemma~\ref{lem:lem1}:} Given any set of real numbers $R_{is}$, $R_{id}$,  $i \in \N$ giving $\wbar$ in \eqref{eq:wbar}, we will prove the lemma by establishing a number of properties for the these numbers in terms of $\wbar$. These properties naturally suggest an algorithm to discover a subset $\Gamma\in \N$ such that $|\Gamma|\leq k$ and $\w(\g) \geq \frac{k}{k+1} \wbar$.

Given any set of real numbers $R_{is}$, $R_{id}$,  $i \in \N$, we have the following property
\begin{itemize}
 \item Property (1): $\exists p \in \n$ such that $R_{ps} \geq \frac{k}{k+1} \wbar$ and $R_{pd} \geq \frac{1}{k+1} \wbar. $ 
If not, we would have the following contradictory argument: Assume for all $i\in\N$, we either have $R_{is} < \frac{k}{k+1} \wbar$ or  $R_{id} < \frac{1}{k+1} \wbar$. Let $S=\{i:R_{is} \geq \frac{k}{k+1} \wbar\}$. By the assumption, this means that $\forall i\in S$, $R_{yd} < \frac{1}{k+1} \wbar$. Therefore considering the subset $S\subseteq\N$, we can upper bound $\wbar$ as,
\begin{align*}
  \wbar  &\leq \max_{i\in S} R_{id} + \max_{i\in \overline{S}} R_{is} \\
      &<\frac{1}{k+1} \wbar + \frac{k}{k+1} \wbar = \wbar.
 \end{align*}
which is a contradiction.

\item Case 1: $R_{pd} \geq \frac{k}{k+1} \wbar$. In this case, the lemma is proved since $\w(\{p\})=\min \lp R_{ps}, R_{pd} \rp \geq  \frac{k}{k+1} \wbar$, and therefore $\w_k\geq\w_1\geq\frac{k}{k+1}\wbar$. 
\end{itemize}
Note the proof is complete for $k=1$ at this point, since $R_{pd} \geq \frac{k}{k+1} \wbar$ is necessarily the case. We assume that $k>1$ in the remaining discussion.

\begin{itemize}
\item Case 2: $R_{pd} < \frac{k}{k+1} \wbar$. Then we have the following property.

Property (2):  $\exists m \in \n$, $m\neq p$ such that $R_{ms} \geq \frac{1}{k+1} \wbar$ and $R_{md} \geq \frac{k}{k+1} \wbar$. Otherwise, we would have the following contradiction: Assume for all $i\in \n$, $i\neq p$,  we either have $R_{is} < \frac{1}{k+1} \wbar$ or  $R_{id} < \frac{k}{k+1} \wbar$. Let $S=\{i\in\n:R_{is} \geq \frac{1}{k+1} \wbar\}$. By Property (1) above, $p\in S$. Moreover, $\forall i\in S$, $R_{id} < \frac{k}{k+1} \wbar$. For $p$ this follows since we are in Case 2 and for other $i\in S$ it follows by the assumption. Therefore  we can upper bound $\wbar$ by
\begin{align*}
  \wbar  &\leq \max_{i\in S} R_{id} + \max_{i\in \bar{S}} R_{is} \\
      &<\frac{k}{k+1} \wbar + \frac{1}{k+1} \wbar = \wbar
 \end{align*}
which is a contradiction.
\end{itemize}

Without loss of generality we can rearrange $i \in \n$ and assume that $p=N$, i.e., $R_{Ns} \geq \frac{k}{k+1} \wbar$ 
and $\frac{k}{k+1} \wbar > R_{Nd} \geq \frac{1}{k+1} \wbar$. Equivalently, $$R_{Ns} \geq \frac{k}{k+1} \wbar\quad\text{and}\quad\frac{k-a+1}{k+1} \wbar > R_{Nd} \geq \frac{k-a}{k+1} \wbar,$$ for an integer $a$ such that $1 \leq a \leq k-1$. Similarly, we can also assume that $m=1$, i.e., $ R_{1s} \geq \frac{1}{k+1} \wbar$ and $R_{1d} \geq \frac{k}{k+1} \wbar$. We proceed by investigating two possible case for $R_{1s}$.

\begin{itemize}
\item Case 1: $R_{1s} \geq \frac{a}{k+1} \wbar$. In this case, the lemma is proved since we would have 
$$\w(\{1,N\}) > \frac{k}{k+1} \wbar,$$ 
which means $w_k\geq w_1\geq\frac{k}{k+1} \wbar$. 
\end{itemize}
Note that the proof is complete for $k=2$ at this point, since $1 \leq a \leq k-1$ yields $a=1$ and
$R_{1s} \geq \frac{a}{k+1} \wbar$ is necessarily the case. We assume that $k>2$ in the remaining discussion.

\begin{itemize}
\item Case 2: $\frac{a}{k+1} \wbar > R_{1s} \geq  \frac{1}{k+1} \wbar$. Equivalently, $$\frac{a_1+1}{k+1}\wbar > R_{1s} \geq \frac{a_1}{k+1} \wbar\quad\text{and}\quad R_{1d} \geq \frac{k-a_0}{k+1} \wbar,$$ for integers $a_1$ and $a_0$ such that $1\leq a_1<a$ and $a_0=0$.

We investigate this case, by proving the following proposition.

\end{itemize}

\begin{proposition}\label{prop} Given positive real numbers $R_{is}$, $R_{id}$,  $i \in \N$, assume that we can arrange them in the following form.
\begin{itemize}
\item $R_{Ns} \geq \frac{k}{k+1} \wbar$ and $\frac{k-a+1}{k+1} \wbar > R_{Nd} \geq \frac{k-a}{k+1} \wbar$ for some  $a \in \mathbb{N}$ such that $1 \leq a \leq k-1$.
\item For any $r$  such that $1 \leq r \leq l$, $\frac{a_r+1}{k+1}\wbar > R_{rs} \geq \frac{a_r}{k+1} \wbar$ and $R_{rd} \geq \frac{k-a_{r-1}}{k+1} \wbar$ for some  $l\in \mathbb{N}$, $1\leq l\leq k-2$, and $a_0, a_1, \dots, a_{l}\in \mathbb{N}$ such that  $a_0=0<a_1<\dots<a_{l-1}<a_{l}<a$.
\end{itemize}
Then, there exists a $y \in \set{l+1,N-1}$ such that $ R_{ys} \geq \frac{a_{l}+1}{k+1} \wbar$ and $R_{yd} \geq \frac{k-a_{l}}{k+1} \wbar$.
\end{proposition}

Before proving the proposition, we first use it to complete the proof of Lemma~\ref{lem:lem1}. Note that we have currently proven that for any positive real numbers $R_{is}$, $R_{id}$,  $i \in \N$, either $r_k\geq \frac{k}{k+1}$, or the assumptions of the proposition are satisfied for $l=1$. 

Assume that the assumptions of the proposition are satisfied for some $1\leq l\leq k-2$. Then the proposition asserts the existence of
$y \in \set{l+1,N-1}$ such that $ R_{ys} \geq \frac{a_{l}+1}{k+1} \wbar$ and $R_{yd} \geq \frac{k-a_{l}}{k+1} \wbar$ for some $a_{l+1}\in\mathbb{N}$ such that $a_{l}<a_{l+1}<a$. This leads to two possible cases for the newly discovered $y \in \set{l+1,N-1}$:
\begin{itemize}
\item Case 1: $R_{ys} \geq \frac{a}{k+1} \wbar$. In this case, the proof of the lemma is completed, because $$\w(\set{l}\cup\{y,N\}) \geq \frac{k}{k+1} \wbar,$$  and $|\set{l}\cup\{y,N\}|\leq k$. This can be observed as follows: Assume  $R_{ys} \geq \frac{a}{k+1} \wbar$ and $R_{yd} \geq \frac{k-a_l}{k+1} \wbar$ for some $y \in \set{l+1,N-1}$. Note that if $\w(\set{l}\cup\{y,N\}) < \frac{k}{k+1} \wbar,$ there exists at least one set $S\subseteq \set{l}\cup\{y,N\}$ such that 
\begin{equation}\label{eq:S}
\lp \max_{i \in S} R_{id} + \max_{i \in \set{l}\cup\{y,N\}\setminus S} R_{is} \rp<\frac{k}{k+1} \wbar.
\end{equation} 
We argue below that such a set $S$ does not exist. Since $R_{Ns} \geq \frac{k}{k+1} \wbar$ we should have $N \in S$. Then also $y\in S$, since otherwise we get the contradiction,
\begin{align*}
\max_{i \in S} R_{id} + \max_{i \in  \set{l}\cup\{y,N\}\setminus S} R_{is} &\geq R_{Nd} + R_{ys} \\
 &\geq \frac{k-a}{k+1} \wbar + \frac{a}{k+1} \wbar \\
&= \frac{k}{k+1} \wbar.
\end{align*}
Then by the same reasoning, we also have $l\in S$. Otherwise,
\begin{align*}
\max_{i \in S} R_{id} + \max_{i \in  \set{l}\cup\{y,N\}\setminus S} R_{is} &\geq R_{yd} + R_{ls} \\
 &\geq \frac{k-a_l}{k+1} \wbar + \frac{a_l}{k+1} \wbar\\
& = \frac{k}{k+1} \wbar.
\end{align*}
Similarly for every $r \in \set{l-1}$, we should also have $r \in S$. This is because if $r+1 \in S$ and $r \in  \set{l}\cup\{y,N\}\setminus S $ we have the following contradiction,
\begin{align*}
\max_{i \in S} R_{id} + \max_{i \in  \set{l}\cup\{y,N\}\setminus S} R_{is} &\geq R_{r+1,d} + R_{rs} \\
 &\geq \frac{k-a_r}{k+1} \wbar + \frac{a_r}{k+1} \wbar \\
&= \frac{k}{k+1} \wbar.
\end{align*}
Therefore $S= \set{l}\cup\{y,N\}$. However, then we have
$$
\max_{i \in S} R_{id}\geq \frac{k-a_0}{k+1}\wbar,
$$
which contradicts \eqref{eq:S} since $a_0=0$.

\item Case 2: $\frac{a}{k+1}\wbar >R_{ys}\geq \frac{a_{l}+1}{k+1}\wbar $. Without loss of generality we can rearrange $y\in \set{l+1,N-1}$ and assume that  $ \frac{a}{k+1} \wbar> R_{l+1,s} \geq \frac{a_l+1}{k+1} \wbar$ and $R_{l+1,d} \geq \frac{k-a_l}{k+1} \wbar$. Equivalently, $$
\hspace*{-0.3cm}\frac{a_{l+1}+1}{k+1}\wbar > R_{l+1,s} \geq \frac{a_{l+1}}{k+1} \wbar\quad\text{and}\quad R_{l+1,d} \geq \frac{k-a_l}{k+1} \wbar,$$ for some $a_{l+1}\in\mathbb{N}$ such that $a_{l}<a_{l+1}<a$. Therefore, we have proven that the assumptions of the proposition should indeed be satisfied with $l+1$ in this case.

\end{itemize}

This implies that starting with $l=1$, we can apply the proposition recursively as long as $l\leq k-2$. At each step of the recursion, either we prove that $r_k\geq \frac{k}{k+1}\wbar$ and the proof of the lemma is complete or $l$ is increased by $1$. Assume that $l=k-2$ and applying the proposition still does not prove the lemma (i.e., the $k$-relays discovered do not satisfy $w(\Gamma)\geq \frac{k}{k+1}\wbar$). Then the proposition establishes the existence of a sequence of positive numbers $ a_0, a_1, a_2, \cdots, a_{k-1}$ such that 
\[
a_0=0<a_1<\dots<a_{k-2}<a_{k-1}<a\leq k-1,
\] 
which is a contradiction. This implies that Case 1 should have been true in one of the earlier iterations of the proposition, which proves the lemma. 

To summarize the conclusions from Case 1 in the above discussion, we have shown that given any positive real numbers $R_{is}$, $R_{id}$,  $i \in \N$ and $1\leq k<N$, they can be either arranged as 
$$
R_{Ns}\geq \frac{k}{k+1} \wbar \quad\textrm{and}\quad R_{Nd}\geq \frac{k}{k+1} \wbar,
$$
or
\begin{itemize}
\item $R_{Ns} \geq \frac{k}{k+1} \wbar$ and $\frac{k-a+1}{k+1} \wbar > R_{Nd} \geq \frac{k-a}{k+1} \wbar$ for some  $a \in \mathbb{N}$ such that $1 \leq a \leq k-1$,
\item and for $1 \leq r \leq l$, $\frac{a_r+1}{k+1}\wbar > R_{rs} \geq \frac{a_r}{k+1} \wbar$ and $R_{rd} \geq \frac{k-a_{r-1}}{k+1} \wbar$ for some  $l\in \mathbb{N}$ such that $1\leq l\leq k-2$, and $a_0, a_1, \dots, a_{l}\in \mathbb{N}$ such that  $a_0=0<a_1<\dots<a_{l-1}<a_{l}<a$,
\item and $R_{l+1,d} \geq \frac{a}{k+1} \wbar$ and $R_{yd} \geq \frac{k-a_{l}}{k+1} \wbar$.
\end{itemize}
For these $l+2\leq k$ nodes $\Gamma=[l+1]\cup \{N\}$, we have $w(\Gamma)\geq \frac{k}{k+1}\wbar$.
 $\hfill\square$

\textit{Proof of Proposition~\ref{prop}:} If the proposition were not true, then we would have the following contradictory argument: Assume for all $i\in\set{l+1,N-1}$, we either have $R_{is} < \frac{a_l+1}{k+1} \wbar$ or  $R_{id} < \frac{k-a_l}{k+1} \wbar$. Let $S=\{i\in\set{l+1,N-1}:R_{is} \geq \frac{a_l+1}{k+1} \wbar\}$. This means that $\forall i\in S$, $R_{id} < \frac{k-a_l}{k+1} \wbar$ and $\forall i\in \set{l+1,N-1}\setminus S$, $R_{is} < \frac{a_l+1}{k+1} \wbar$. Therefore considering the subset $ S\cup\{N\}\subseteq\N$, we can upper bound $\wbar$ as,
\begin{align*}
  \wbar  &\leq \max_{i\in S\cup\{N\}} R_{id} + \max_{i\in \N\setminus S\setminus\{N\}} R_{is} \\
&= \max_{i\in S\cup\{N\}} R_{id} + \max_{i\in \set{l}\cup(\set{l+1,N-1}\setminus S)} R_{is} \\
      &<\max\lp\frac{k-a_l}{k+1} \wbar,\, \frac{k-a+1}{k+1} \wbar\rp + \max_{1\leq r\leq l}\frac{a_r +1}{k+1} \wbar \\
&= \frac{k-a_l}{k+1} \wbar+ \frac{a_l +1}{k+1}\wbar =\wbar,
 \end{align*}
which is a contradiction.  $\hfill\square$

\bigskip
\textit{Proof of Lemma~\ref{lem:lem2}:} We will prove that for the configuration $R_{is} = i\,R$ and $R_{id} = (k+2-i)\,R$ for $1\leq i\leq k+1$, we have $\w_k=\frac{k}{k+1}\wbar$.

We first show that for this particular configuration $\wbar= (k+1)R$. Let $\l$ be any subset of $\set{k+1}$ and let $y(\l)=\max_{i \in \bar{\l}} R_{is}$. Then, $\max_{i \in \l} R_{id} \geq (k+2)R - \lp y(\l)+R \rp$. Note that the last inequality holds even if $y(\l)=(k+1)R$. Therefore, we have
 \begin{align*}
  \wbar &= \min_{\l \subseteq \set{k+1}} \lp \max_{i \in \l} R_{id} + \max_{i \in \lbar} R_{is} \rp \\
        &\geq \min_{\l \subseteq \set{k+1}} \lb \lp k+1-y(\l) \rp + y(\l) \rb = (k+1)R.
 \end{align*}
On the other hand, $\wbar\leq (k+1)R$. Therefore, $\wbar=(k+1)R$.

We now prove that for any $\g \subset \set{k+1}$
with $|\g| = k$, we have $\w\lp\g\rp= k R$. Let  $\l$ be any subset of $\g$ and let $y(\l)=\max_{i \in \g\setminus\l} R_{is}$. Then $\max_{i \in \l} R_{id} \geq (k+2)R - \lp y(\l)+2R \rp$. Note that this inequality holds even if $y(\l)=(k+1)R$. The reason that we have used
$y(\l)+2R$ this time is because of the possibility that $\arg\max_{i \in \g\setminus\l} R_{is}+1  \not\in \g$. Therefore, we have,
 \begin{align*}
  \w\lp\g\rp &= \min_{\l \subseteq \g} \lp \max_{i \in \l} R_{id} + \max_{i \in \g\setminus\l} R_{is} \rp \\
        &\geq \min_{\l \subseteq \g} \lb \lp kR-y(\l) \rp + y(\l) \rb = kR.
 \end{align*}
Now, for any $\g \subseteq \set{k+1}$ with $|\g|=k$ there exists a $j\lp\g\rp \in \set{k+1}$ such that $\g = \set{k+1} \setminus \{j\lp\g\rp\}$. Then, we have
 \begin{align*}
  \w\lp\g\rp & = \min_{\l \subseteq \g} \lp \max_{i \in \l} R_{id} + \max_{i \in\g\setminus\l} R_{is} \rp \\
        & \leq \max_{i \in \set{j\lp\g\rp-1}} R_{is} +  \max_{i \in \set{j\lp\g\rp+1,k+1}} R_{id} \\
	& =\lp j\lp\g\rp-1 \rp R+ \lp k+2 - \lp j\lp\g\rp+1 \rp\rp R = kR.
 \end{align*}
Note that this reasoning holds even if $j\lp\g\rp = 1$ or $j\lp\g\rp = k+1$.

Therefore, we have proved that
\begin{equation*}
 \w_k = \max_{\substack{\Gamma \subseteq \set{k+1} \\ |\Gamma|=k}} \w\lp\g\rp = kR.
\end{equation*}

\end{document}

%% file: N-diamond.tex
\begin{center}
\psset{unit=0.014in}
\begin{pspicture}(0,-5)(120,100)
\psset{linewidth=0.5mm}
\definecolor{darkgreen}{rgb}{.15,.35,.15}

\rput(-5,40){\circlenode[fillstyle=solid,fillcolor=lightgray]{S}{{\large $\mathbf{s}$}}}
\rput(60,100){\circlenode[fillstyle=solid,fillcolor=lightgray]{A1}{$1$}}
\rput(60,60){\circlenode[fillstyle=solid,fillcolor=lightgray]{A2}{$2$}}
\rput(60,30){\Large $\vdots$}
\rput(60,-5){\circlenode[fillstyle=solid,fillcolor=lightgray]{AN}{$N$}}
\rput(125,40){\circlenode[fillstyle=solid,fillcolor=lightgray]{D}{$\mathbf{d}$}}

\ncline[linewidth=0.5mm, linestyle=solid]{->}{S}{A1} \Aput{\large ${h_{1s}}$}
\ncline[linewidth=0.5mm, linestyle=solid]{->}{S}{A2} \Bput{\large $h_{2s}$}
\ncline[linewidth=0.5mm, linestyle=solid]{->}{S}{AN} \Bput{\large $h_{Ns}$}

\ncline[linewidth=0.5mm, linestyle=solid]{->}{A1}{D} \Aput{\large $h_{1d}$}
\ncline[linewidth=0.5mm, linestyle=solid]{->}{A2}{D} \Bput{\large $h_{2d}$}
\ncline[linewidth=0.5mm, linestyle=solid]{->}{AN}{D} \Bput{\large $h_{Nd}$}

%
\end{pspicture}
\end{center} 

%% file: diamond.tex
\begin{center}
\psset{unit=0.015in}
\begin{pspicture}(0,-20)(90,90)
\psset{linewidth=0.5mm}
\definecolor{darkgreen}{rgb}{.15,.35,.15}

\rput(-40,70){\bf\large Case (a)}
\rput(0,70){\circlenode[fillstyle=solid,fillcolor=lightgray]{S}{{\large $\mathbf{s}$}}}
\rput(44,92){\circlenode[fillstyle=solid,fillcolor=lightgray]{A}{$1$}}
\rput(44,48){\circlenode[fillstyle=solid,fillcolor=lightgray]{B}{$2$}}
\rput(90,70){\circlenode[fillstyle=solid,fillcolor=lightgray]{D}{\large $\mathbf{d}$}}

\ncline[linewidth=0.5mm, linestyle=solid]{->}{S}{A} \Aput{\large ${t}$}
\ncline[linewidth=0.5mm, linestyle=solid]{->}{S}{B} \Bput{\large $t$}
\ncline[linewidth=0.5mm, linestyle=solid]{->}{A}{D} \Aput{\large $t^2$}
\ncline[linewidth=0.5mm, linestyle=solid]{->}{B}{D} \Bput{\large $t^2$}
%

\rput(-40,0){\bf\large Case (b)}
\rput(0,0){\circlenode[fillstyle=solid,fillcolor=lightgray]{S}{{\large $\mathbf{s}$}}}
\rput(44,22){\circlenode[fillstyle=solid,fillcolor=lightgray]{A}{$1$}}
\rput(44,-22){\circlenode[fillstyle=solid,fillcolor=lightgray]{B}{$2$}}
\rput(90,0){\circlenode[fillstyle=solid,fillcolor=lightgray]{D}{\large $\mathbf{d}$}}

\ncline[linewidth=0.5mm, linestyle=solid]{->}{S}{A} \Aput{\large ${t}$}
\ncline[linewidth=0.5mm, linestyle=solid]{->}{S}{B} \Bput{\large $t^2$}
\ncline[linewidth=0.5mm, linestyle=solid]{->}{A}{D} \Aput{\large $t^2$}
\ncline[linewidth=0.5mm, linestyle=solid]{->}{B}{D} \Bput{\large $t$}
%
\end{pspicture}
\end{center} 